# Privacy-Preserving Hierarchical Anonymization Framework over Encrypted Data

Jing Jia, Kenta Saito, and Hiroaki Nishi, *Member, IEEE*

*Abstract*— Smart cities, which can monitor the real world and provide smart services in a variety of fields, have improved people's living standards as urbanization has accelerated. However, there are security and privacy concerns because smart city applications collect large amounts of privacy-sensitive information from people and their social circles. Anonymization, which generalizes data and reduces data uniqueness is an important step in preserving the privacy of sensitive information. However, anonymization methods frequently require large datasets and rely on untrusted third parties to collect and manage data, particularly in a cloud environment. In this case, private data leakage remains a critical issue, discouraging users from sharing their data and impeding the advancement of smart city services. This problem can be solved if the computational entity can perform the anonymization process without obtaining the original plain text. This study proposed a hierarchical k-anonymization framework using homomorphic encryption and secret sharing composed of two types of domains. Different computing methods are selected flexibly, and two domains are connected hierarchically to obtain higher-level anonymization results in an efficient manner. The experimental results show that connecting two domains can accelerate the anonymization process, indicating that the proposed secure hierarchical architecture is practical and efficient.

## I. Introduction

THE development of smart cities [1], which leverages a variety of promising technologies such as the Internet of Things, information-physical systems, big data analytics, and real-time control, has provided smart services and comfortable living for local residents, as well as influenced people's lives. These systems are not fragmented but interconnected in a collaborative manner. In the structure of a smart city or super city, the data distribution infrastructure (data linking infrastructure) is a critical component [2]. However, the data obtained by the service frequently contains users' personal data, and if these data are not properly managed or used for other purposes, privacy issues such as leakage of users' personal data will arise [3–5]. In fact, a number of issues have arisen in the past because of privacy concerns [6,7].

Data anonymization is the process of protecting private or sensitive information by eliminating or generalizing identifiers that link individuals to the stored data. Anonymized data should not be identifiable by itself, and in some cases, using background information. In contrast to directly deleting sensitive data, anonymization preserves the relevant information for other functional applications. The use of anonymized data in information distribution infrastructure can facilitate the use of data.

Collecting data from multiple data subjects rather than anonymizing data from a single data source produces useful results by reducing information loss. In contrast, collecting large amounts of data requires extensive storage and powerful computational performance for anonymization. Therefore, the original data must typically be outsourced to an external server and processed with the assistance of third parties, such as cloud services. When relying on third parties, the original data are stored on remote servers until the anonymization process is complete. Despite the fact that some regions and institutions have enacted legislation restricting data distribution and retention to facilitate data reuse [8], external retention of pre-anonymized data may also have a negative impact on data utilization. In particular, data owners concerned with making their data available to external parties may be more reluctant to send pre-anonymized data to untrusted third parties [9].

Therefore, in this case, even if pre-anonymized data were temporarily transferred to untrusted third parties, it become a privacy problem and an urgent obstacle to data sharing. This problem can be addressed if the computational entity performing anonymization activities is capable of managing and processing the input data without obtaining plain text. The technology of data anonymization involves sharing the data, and secure computation technology is used to secure data in the computation process of the data. Data anonymization using secure computation is an essential solution for this problem.

Homomorphic encryption is a class of encryption methods that was first proposed by Rivest *et al.* [10] in the 1970s.

Jing Jia is with the Graduate School of Science and Technology, Keio University, Kohoku-ku, Yokohama, 223-8522, Japan (e-mail: jiajing1995@keio.jp).

Kenta Saito was with the Graduate School of Science and Technology, Keio University, Kohoku-ku, Yokohama, 223-8522, Japan (e-mail: saiton15603@gmail.com).

Hiroaki Nishi is with the Faculty of Science and Technology, Keio University, Kohoku-ku, Yokohama, 223-8522, Japan (e-mail: west@sd.keio.ac.jp).

Compared with general encryption algorithms, homomorphic encryption can only be informed of the final result but cannot obtain each plaintext message, thereby improving information security. Therefore, it is feasible to achieve anonymity over the encrypted datasets.

The realization of anonymizing encrypted datasets without exposing sensitive data to untrusted servers can enhance the credibility of infrastructures; however, the processing time also needs to be considered. The concept of secret sharing is to split the secret appropriately, and each share of the split is managed by a different participant [11,12]. The combination of homomorphic encryption and secret sharing can hierarchically divide the anonymization process. We propose a method in which different secret computation methods are applied to different domains for reducing the computational costs.

This study proposes a secure method for anonymization that does not expose the original data to untrusted third parties. Our main contributions are summarized as follows:

(1) We propose a privacy-preserving hierarchical anonymization framework in an encrypted domain using secure computation. The framework focuses on the privacy preservation requirements of data owners, achieves secure anonymization at remote servers, and protects sensitive information.

(2) The proposed hierarchical framework divides the target computational domain into two different domains: local and global, and different secure computational methods are conducted flexibly. More specifically, homomorphic encryption is conducted in the local domain with limited computational power, whereas secret sharing is conducted in the global domain with sufficient computational subjects.

(3) A hierarchical structure is achieved by sending the anonymized results of a local domain to a global domain, which can obtain higher-level anonymization datasets. The experimental results show that the connection of the two domains can accelerate the anonymization process by more than two times.

The remainder of this paper is organized as follows. Section II discusses background information and related literature. In Section III, we provide a detailed description of the construction of the proposed framework. We evaluate the proposed scheme in Section IV. The conclusions are presented in Section V.

## II. Background Information and Related Work

Anonymization is a method of reducing the uniqueness of information and protecting privacy by removing, pseudonymizing, or abstracting portions of information from the data. According to the data privacy terminology, data attributes can be divided into the following types: (1) direct identifiers can directly and uniquely identify an individual, such as name or ID number; (2) quasi-identifiers are those that can be combined with auxiliary information to reveal someone's identity, such as age, gender, and ZIP code; (3) sensitive attributes are information that individuals want to hide from others, such as salary and disease; and (4) attributes other than the above three types are non-sensitive attributes.

One of the anonymization methods is k-anonymity. First proposed by Sweeney *et al*. [13], k-anonymity is defined as "the state in which for any record in a data table, there exist k or more records in the data table with the same combination of quasi-identifiers, including itself." A group of records with the same combination of quasi-identifiers is called a q* block. All q*blocks containing k or more records satisfy k-anonymity. Fig. 1. shows an example of k-anonymization, in which the ZIP code and age of patients are quasi-identifiers, and diseases are sensitive attributes. K-anonymization methods have been developed in recent years [14–17]; however, they do not consider the problem of direct access to unmodified data.

Another technique is differential privacy (DP). DP anony-

|   | ZIP Code | Age | Disease       |   | ZIP Code | Age  | Disease       |
|---|----------|-----|---------------|---|----------|------|---------------|
| 1 | 47677    | 29  | Heart Disease | 1 | 476**    | 2*   | Heart Disease |
| 2 | 47602    | 22  | Heart Disease | 2 | 476**    | 2*   | Heart Disease |
| 3 | 47678    | 27  | Heart Disease | 3 | 476**    | 2*   | Heart Disease |
| 4 | 47905    | 43  | Flu           | 4 | 4790*    | ≥40  | Flu           |
| 5 | 47909    | 52  | Heart Disease | 5 | 4790*    | ≥40  | Heart Disease |
| 6 | 47906    | 47  | Cancer        | 6 | 4790*    | ≥40  | Cancer        |
| 7 | 47605    | 30  | Heart Disease | 7 | 476**    | 3*   | Heart Disease |
| 8 | 47673    | 36  | Cancer        | 8 | 476**    | 3*   | Cancer        |
| 9 | 47607    | 32  | Cancer        | 9 | 476**    | 3*   | Cancer        |

(a) original patients table   (b) k-anonymized table (k = 3)

Fig. 1. Example of K-anonymization

mizes data by inserting Laplace noise into the data when using the Laplace mechanism. The process of inserting noise can be easily replaced with secret computation, which has already been proposed and achieved by the authors, and is not mentioned here.

Homomorphic encryption is a feasible solution for achieving operations over encrypted data. It can achieve both basic encryption operations and multiple computation functions of addition and multiplication between ciphertexts [18]. Additive homomorphic encryption is expressed by the following equation:

$$Dec(Enc(x_1) \oplus Enc(x_2)) = x_1 + x_2 \quad (1)$$

where $\oplus$ represents the addition operator between ciphertexts, and $Enc(\cdot)$ and $Dec(\cdot)$ denote the encryption function and the decryption function, respectively. Similarly, multiplicative homomorphic encryption is expressed by the following equation:

$$Dec(Enc(x_1) \otimes Enc(x_2)) = x_1 \times x_2 \quad (2)$$

where $\otimes$ represents the multiplication operator between ciphertexts.

Considering the trade-off between privacy and functionality, a distributed framework to realize secure anonymization was developed. Jiang *et al*. [19] proposed a two-party secure distributed framework that could jointly execute k-anonymization from two vertically partitioned sources, without disclosing data from one site to other. Each party performed local generalization until the data were sufficiently anonymized using commutative cryptography and a secure set intersection protocol. The framework proposed in [19] was a two-party semi-honest framework, and we consider constructing a hierarchical framework using different secure computation methods for multiple parties to resist malicious attacks.

Secret sharing [11,12] is an appropriate method to distribute computational tasks to different participants. A single participant cannot recover the secret message, and only several participants working together can recover it. More importantly, when something goes wrong with any of the corresponding ranges of participants, the secret can still be recovered.

## III. Proposed Scheme

### A. Overview

This study assumes an environment in which computational resources are arranged in an hierarchical manner and secret computations are used to perform k-anonymization processing based on k-member clustering. Fig. 2. shows a systematic overview of the environment envisioned in this study. The topology of the network is hierarchical and computational resources for anonymization services exist at each level. The assumed environment includes the following areas and entities:

(1) Global domain: A network domain in which there are several computational entities that can compute anonymization processes within the domain and in which there is sufficient space to distribute shares when secret sharing is used. For example, a cloud domain can be treated as a global domain.

(2) Local domain: A network domain in which there are few computational entities that can compute anonymization pro-

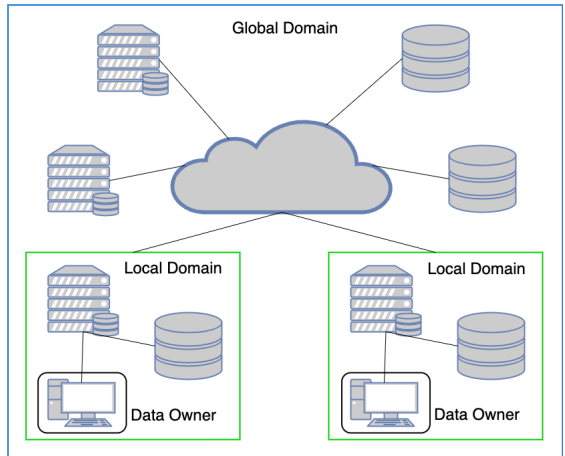

Fig. 2. Overview

cesses within the domain and where there are insufficient entities to distribute shares when secret sharing is used. The edge domain corresponds to this domain.

(3) Data Owners: Data owners possess confidential information and provide data for various services. Data owners enter into a contract in advance with a service provider that stipulates the type of data to be provided, type of anonymization, and privacy standards for anonymization results.

(4) Computation Units: Computation units provide large storage and assist data owners process anonymization of encrypted datasets.

In this study, the anonymization process was hierarchically divided. Different secret computation methods were applied to each domain.

In the local domain, the number of computational resources is small, and there are insufficient distributed locations for computation when secret sharing is employed. The required computational resources are not very large, because the amount of data is similar not so large. Therefore, homomorphic cryptography is used for secret computations. However, in the global domain, the number of computational operators is large and widely spread. In this domain, because the data are gathered and their size increases, the required computational power also increases significantly. Secret sharing, which requires less processing time per operation than homomorphic cryptography, can reduce the time and cost, and is more feasible because the amount of data is larger than that in the local domain. Thus, the two secret computations were used together to achieve secure k-anonymization and reduce the overall computational cost.

The degree of generalization is based on the number of computational resources in a domain. Therefore, it is assumed that the local and global domain anonymization processes can be linked. Consequently, anonymization results from the local domain are collected in the global domain and anonymized more robustly than anonymized data in the local domain. This allows each service provider to use data that meet the appropriate privacy criteria. In addition, the processing time can be reduced compared with the time required for anonymization in the global domain.

The rules for separating the global domain from the local domain depend on the data to be anonymized and the number of data owners of the computing resources. These rules must be defined in advance. An example of this is as follows: Let $N_D$ be the number of owners of the computing resources that can be anonymized in the domain. In this case, the following criteria were used to select the secret computation method:

(1) homomorphic encryption is used when
$N_D = 1$;
$N_D \geq 2$ when most of the calculated resources are possessed by the same entity.

(2) Secret sharing is used in cases other than those described above.

When $N_D = 1$, it is impractical to distribute to multiple computing entities with secret sharing, and even when $N_D \geq 2$, most of the calculated resources are possessed by the same entity. It is inappropriate to apply secret sharing in terms of resistance to collusion attacks. Therefore, the criteria for choosing which secure computation method to use is defined above.

### B. Dataflow

This subsection describes the data flow in the proposed anonymization process. A sequence diagram of a data owner and computation subject using homomorphic encryption is shown in Fig. 3. This sequence diagram omits the process when it fails. When the local and global domains are linked, the computational entity of the local domain becomes the data owner of the global domain.

The data flow of the anonymization process is shown in Fig. 3, corresponding to the following numbers:

1. The data owner sends data records to the computation unit.
2. The computation unit determines the secret computation.
3. The computation unit sends the secret computation context to the data owner.
4. The data owner performs key generation, etc.

5. The data owner encrypts confidential information.

6. The data owner sends ciphertext and a public key to the computation unit.

7. The computational unit performs the anonymization process. During this process, the computational unit may request computational support several times and use the results of the computational support for the anonymization process.

8. The computation unit sends the anonymization result (ciphertext) to the data owner.

9. The data owner decrypts the anonymization result.

10. The data owner sends the anonymization result (plaintext) to the computation unit.

11. The computation unit registers the anonymization result in a database.

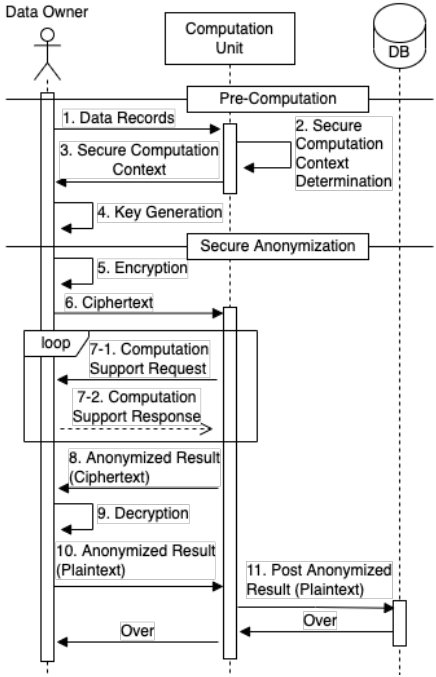

Fig. 3. Data flow (homomorphic encryption)

Fig. 4. shows the sequence diagram of the anonymization process using secret sharing. Compared with anonymization using homomorphic encryption, there is a controller that manages secret sharing computation. The data owner sends the data records to the controller to generate a secret computation context. The secret computation context includes an ID that identifies the secret computation process and destination of the share. Then, the data owner sends the secret computation context to each computation unit, and the following flow is the same as that using homomorphic encryption.

### C. K-anonymization

Hereafter, the number of records in a dataset is denoted as $N$, and the number of quasi-identifiers is denoted as $M$. In the anonymization process, identifiers are deleted and quasi-identifiers are transformed into a range. Records in the same q* block have the same quasi-identifiers as in the definition of k-anonymity.

#### 1) K-member clustering

First, the data owner transforms and normalizes the numerical data values of the quasi-identifiers such that the original data values fall within the range of [0,1]. Specifically, the value $x$ is converted to $x'$ and normalized using the following formula:

$$x' = \frac{x - x_{min}}{x_{max} - x_{min}} \quad (3)$$

where $x_{max}$ represents the maximum value of this dataset and $x_{min}$ represents the minimum value of this dataset.

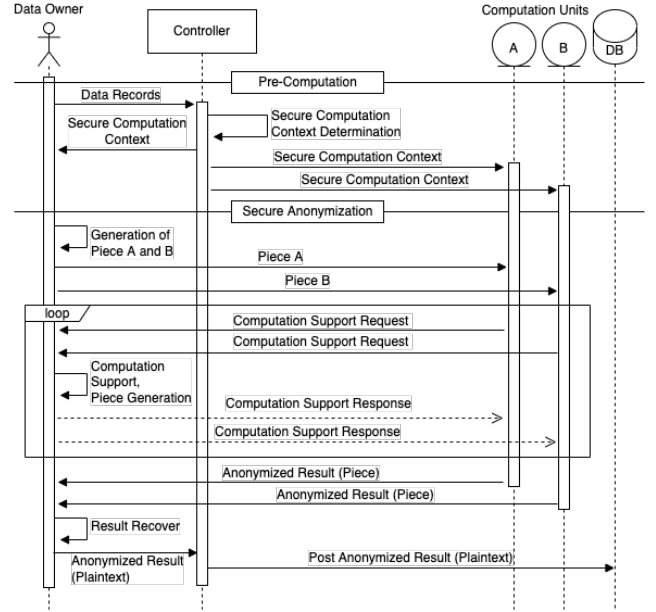

Fig. 4. Data flow (secret sharing)

The data owner encrypts the normalized data and sends them to the computation unit. The computational unit performs k-member clustering using the received data (ciphertext). Fig. 5. shows the pseudo-code for k-member clustering in the proposed method. First, the order of records is randomly determined. Subsequently, based on the order, if the chosen record is unclassified, it is selected as a core record and added to the cluster. A record that is close to the core record is then selected and added to the cluster to generate a cluster containing k records after k-1 rounds. The maximum number of clusters that can be created when N records are classified to satisfy anonymity is $\lfloor N/k \rfloor$ records. Therefore, once $\lfloor N/k \rfloor$ clusters are created, the creation of new clusters is stopped. Then, unclassified records must identify the closest classified record and add to the cluster to which they belong.

#### 2) Find the closest record

The target records must be selected to be added to the cluster. This process corresponds to the find_best_record_index function shown in Fig. 6. First, the distance between the core record and each candidate record is calculated. The distance is calculated as the square of the difference between the records for each quasi-identifier and the sum of the squares. In other words, the distance $dist(u, v)$ between records $u$ and $v$ is defined as follows:

$$dist(u, v) = \sum_{i=0,\dots,M-1} (a_{u,i} - a_{v,i})^2 \quad (4)$$

In the fourth line of Fig. 6, the distance is divided by $M$ (multiplied by $1/M$) because it is necessary to set $0 \le d_i \le 1$ ($0 \le i \le M - 1$) for the convenience of the $max\_index$ in the ninth line.

Because the secret computation encrypts the result of the distance between the core record and each record, we do not know the smallest value. Therefore, we used a calculation aid to obtain the minimum array value. However, if an array of distances is given to the private key holder for computational support, the original data may be exposed. $max\_index$ uses the following formula to convert the distances from $x_i$ to $y_i$. The more iterations $t$ is set, the closer the value approaches 0 and 1.

$$y_i = \frac{x_i^{2^t}}{\sum_j^N x_j^{2^t}} \sim \begin{cases} 1 & (x_i = \max x) \\ 0 & (\text{otherwise}) \end{cases} \quad (5)$$

When the value of the distance is $0 \le d_i \le 1$ ($0 \le i \le M-1$), the index of the record with the smallest distance can be obtained by computing $x_i = 1 - d_i$. Because find_best_record _index extracts the closest records from a single core record, the distance to each record is calculated only once. Subsequently, $max\_index$ is executed with a mask applied to the classified record to prevent it from being selected as the record with the smallest distance, and the result is repeated to obtain the result with calculation support. This allowed us to extract the nearest records from the core records.

*3) Generate output data*

First, the range of quasi-identifiers in each q* block is determined to generate the output data, and confidential information is transformed into a list form. First, we explain how to determine the range of quasi-identifiers in a q* block. Because the number of q* blocks is $\lfloor N/k \rfloor$, the number of values subject to computation support is $2M\lfloor N/k \rfloor$. Because the result of the computation, support depends on the result of each computation support, whereas the computation support in output data generation can calculate all values independently, only one communication is required even if the computation support must be performed in sequence. The method for converting the confidentiality information into a list format is to compile the confidentiality information of all records in the q* block into a list format. Subsequently, the output data of the anonymization process were generated.

*4) Decryption and recovery of anonymized results*

Because all calculation results from secret calculations are encrypted, it is necessary to decrypt the anonymized results. Therefore, as shown in Fig. 3, in steps 8 to 10, the anonymization results in encrypted form are sent to the data owner for decryption. However, if the data owner sends fake records back to the computation unit, it is unaware that the data owner has tampered with the result. This section discusses the proposed methods for detecting tampering during decryption.

In homomorphic cryptography, the computation unit can probabilistically detect tampering by adding dummy data to the anonymized cipher result. In this case, instead of combining q* blocks into a single record, a dummy record was generated for each record in the original data table. The encryption method used for homomorphic encryption was a public-key cryptosystem; therefore, there were no security problems even if the public key used for encryption was made public. The computation unit generated encrypted dummy data and inserted it into the anonymization result before the decryption request and passed it to the private key holder. If the value of the dummy data was rewritten, it was certain that the data had been tampered. The random numbers added to the data records also prevented the data owner from guessing which data were dummy data from the decryption results. If only some of the records in a q* block had been tampered, k-anonymity was no longer satisfied when noise was removed from the decryption results and tampering could be detected.

For secret sharing, results can be restored if a certain number of shares are collected. Therefore, multiple computation units and data owners can share pieces after the anonymization

---

**Function k_member_clustering**(*S*, *k*)

Input: Array of data records *S*, anonymization parameter *k*
Output: Clusters each of which contains at least *k* records
1. Initialize array *clusters*
2. *cluster_count*s = floor (*S*, *k*)
3. *N* = len(*S*)
4. Initialize array *used* with the number of records *N* and value *false*
5. *rand_array* = [0, .., *N*-1]
6. Shuffle *rand_array*
7. Initialize array *rests*
8. for i, v in *rand_array*:
9.    if *used*[v]:
10.      continue
11.    *used*[v] = *true*
12.    Initialize array *c* and insert v
13.    *record_list* = **find_best_record_index** (*S*, *used*, v, *k*-1)
14.    Insert all elements of *record_list* to *c*
15.    Insert *c* into *clusters*
16.    if len(*clusters*) == *cluster_counts*:
17.      *rests* = *rand_array*[i+1:]
18.      break
19. for v in *rests*:
20.    if *used*[v]:
21.      continue
22.    *used*[v] = *true*
23.    *reverse_used* = reversed value of *used*
24.    r = **find_best_record_index**(*S*, *reverse_used*, *c*, 1)
25.    *cluster* = cluster containing *r*
26.    Insert v into *cluster*
27. return clusters

Fig. 5. Pseudo code of **k_member_clustering**

---

**Function find_best_record_index** (*S*, *used*, *r*, *count*)

Input: Array of encrypted data records *S*, clustered flag array of records *used*, encrypted core record *r*, number of records to extract *count*
Output: Index of nearest *count* records to *r*
1. Initialize array *d* with size *N* and value 0
2. for i, v in *S*:
3.    for m in range(*M*):
4.      *d*[i] = *d*[i] +$_{enc}$ (*dist*(r[m], v[m]) ×$_{enc}$ $\frac{1}{M}$)
5.    *d*[i] = Enc(1) −$_{enc}$ *d*[i]
6.    If *used*[i] == *true*, then *d*[i] = 0
7. Initialize array *top_index*
8. for i in range(*count*):
9.    *best_index* = *max_index*(*d*)
10.    Insert *best_index* into *top_index*
11.    *used*[*best_index*] = *true*, *d*[*best_index*] = 0
12. return *top_index*

Fig. 6. Pseudo Code of **find_best_record_index**

process, enabling multiple people to recover the results. Each entity then sends its decryption results to the controller, which checks them against each other to determine if they have been tampered.

## IV. Experimental Results

### A. Evaluation Criterion

The evaluation criteria are the execution time of the anonymization process and the information loss of the anonymization result. The following formula is used to calculate the information loss, where the number of records is denoted as $N$, and the number of quasi-identifier attributes is denoted as $M$.

$$IL = \frac{1}{N}\sum_{i=1,\dots,N}\sum_{j=1,\dots,M}\frac{a_{i,j,max}-a_{i,j,min}}{|N_j|} \qquad (6)$$

where $a_{i,j,max}$ and $a_{i,j,min}$ represent the upper and lower limits of the range values of the *i*-th record and *j*-th quasi-identifier, respectively. $|N_j|$ represents the domain size of the *j*-th quasi-identifier.

### B. Execution Time

In this section, we describe the case in which only homo-morphic cryptography (local domain) and secret sharing is used (global domain). The execution time was measured from the time when the encryption of the data table values began after key generation to the time when the anonymization process ended after decryption.

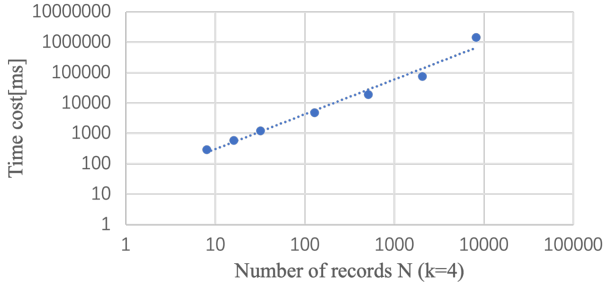

Fig. 7. Execution time with different N
(using homomorphic encryption)

#### 1) Homomorphic Encryption

First, we evaluated the case in which only homomorphic cryptography was used. Fig. 7 shows the execution time for a different number of records when k = 4 is fixed, and Fig. 8 shows the execution time for a different anonymization parameter k when N = 512 is fixed.

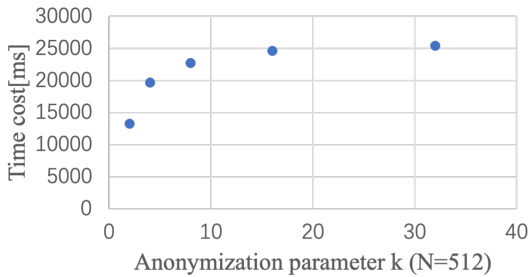

Fig. 8. Execution time with different k
(using homomorphic encryption)

The execution time increased linearly with the number of records. In the anonymization process, one record was randomly selected from the unclassified records as the core record, and the process of determining a record that is close to the core record was executed $k-1$ times. The number of times a core record is selected is equal to the number of q* blocks and is $[N/k]$ times, which is approximately $N/k$ times. Therefore, because each q* block generation involves k-1 record selection operations and generates q* blocks approximately N/k times, the time complexity of the anonymization process can be expressed using the following equation:

$$O\left(\frac{N}{k}\cdot(k-1)\right) = O\left(N-\frac{N}{k}\right) \qquad (7)$$

Regarding the increase in execution time with different k values, it can be observed that the execution time increases as k increases; however, the increase in execution time becomes milder. This can be explained by the fact that the time complexity of the anonymization process can be expressed by (7).

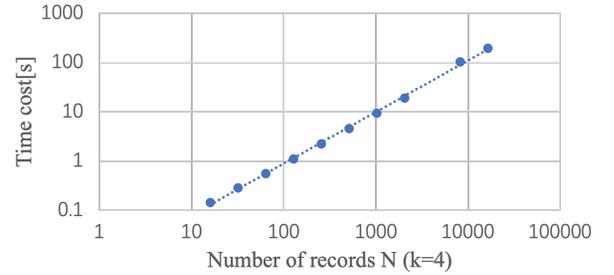

Fig. 9. Execution time with different N
(using secret sharing)

#### 2) Secret Sharing

Fig. 9 shows the change in execution time for the anonymization process using secret sharing under different values of

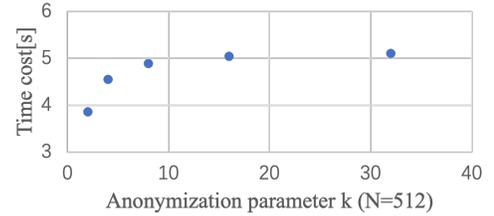

Fig. 10. Execution time with different k
(using secret sharing)

N when k = 4 is fixed, and Fig. 10 shows the change in execu-tion time under different values of k when N = 512 is fixed. It can be observed that the execution time increases with O(N-N/k), as in the case of homomorphic encryption. Compared to the homomorphic encryption method, the secret-sharing method is generally 3.9 to 4.3 times faster than the homomorphic encryption method.

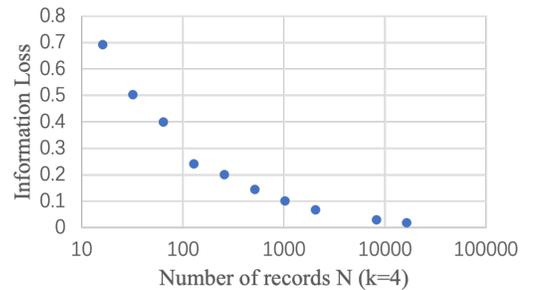

Fig. 11. Information loss with different N

### 3) Information Loss

Fig. 11 shows the information loss with a different number of records when k = 4, and Fig. 12 shows the information loss with a different value of k when N = 512. The information loss with the number of records shows that the information loss decreased as the number of records increased and the number of requests increased.

### 4) Hierarchical Connection of Two Domains

As mentioned, the experiments were conducted assuming an environment in which anonymization processes were coordi-nated in the local domain as towns and global domains as cities. The following is a description of the assumed experimental environment. Assuming that there were eight local domains that did not have sufficient share distribution destinations, homomorphic encryption was used to generate k = 4 anonymity, and there were sufficient operators in global domains that could perform secret sharing to satisfy k =16 anonymity.

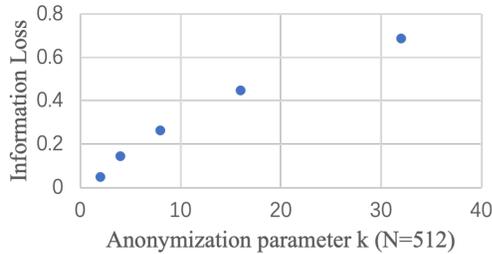

Fig. 12. Information loss with different k

As shown in Table 1, when using homomorphic encryption in the local domain and secret sharing in the global domain, the total execution time for our proposed hierarchical scheme was $1.19 \times 10^2$ s by sending the anonymized results of local do-mains to the global domain as input. Conversely, the process took $1.53 \times 10^2$ s when using homomorphic cryptography in both domains, indicating that secret sharing accelerated the process. In addition, when the results of the local domain were not used in the global domain, the process took $3.11 \times 10^2$ s, indicating that the connection of the local and global domains sped up the anonymization process more than two times.

TABLE I
TIME COST

| Global Domain | Homomorphic Encryption | Secret Sharing | Secret Sharing |
|---|---|---|---|
| Local Domain | Homomorphic Encryption | Homomorphic Encryption | Homomorphic Encryption |
| Connection | ○ | ○ | × |
| Time (s) | $1.53 \times 10^2$ | $1.19 \times 10^2$ | $3.11 \times 10^2$ |

Moreover, when homomorphic encryption was used in the local domain and secret sharing was used in the global domain, the information loss per record was 0.113 if the global domain used the anonymized results of the local domain, and the information loss per record without using the anonymized results of the local domain was 0.107. The increase in information loss owing to the connection between the two domains was 5.6%. However, the processing time decreased by 61.7% owing to this connection. Therefore, in practical use, an increase in information loss is acceptable.

## V. CONCLUSION

Privacy-preserving outsourcing and data sharing will become increasingly important with the development of smart cities. However, data sharing may cause leakage of sensitive information, and existing anonymization methods that rely on untrusted third parties still expose unmodified data.

In this study, we propose a privacy-preserving hierarchical framework for k-anonymization using two secret computation methods: homomorphic encryption and secret sharing. The network was divided into global and local domains. The experimental results showed that the execution time increased linearly with the number of records and could be reduced by sending the anonymized results of a local domain to a global domain to realize hierarchical anonymization. More specifically, using secret sharing in the global domain could deduce over 20% of the process time, and the connection between the local and global domains could reduce the process time by 61.7 %, whereas the information loss only increased by 5.6%. Hence, the framework proposed in this study can achieve secure anonymization of encrypted data and speed up the process, benefiting from the hierarchical structure. The proposed methods will be imported into our testing smart city data platform in Saitama City, Japan.

## REFERENCES

[1] IBM Institute for Business Value, "A vision of smarter cities," 2009. [Online]. Available: https://www.ibm.com/downloads/cas/2JYLM4ZA.
[2] Secretariat for Promotion of Regional Development, Cabinet Office, Government of Japan, "About super cities and digital rural health special zones," 2021. [Online]. Available: https://www.chisou.go.jp/tiiki/kokusentoc/supercity/openlabo/supercity.pdf.
[3] K.D. Martin, A. Borah, and R.W. Palmatier, "The dark side of big data's effect on firm performance," Marketing Science Institute Working Paper Series, Rep. 16–104, Jan. 2016.
[4] K. Adhikari and R. K. Panda, "Users' information privacy concerns and privacy protection behaviors in social networks," J. Glob. Market., vol. 31, no. 2, pp. 96–110, Jan. 2018, DOI: 10.1080/08911762.2017.1412552, [Online].
[5] J. Wieringa, P. K. Kannan, X. Ma, T. Reutterer, H. Risselada and B. Skiera, "Data analytics in a privacy-concerned world," J. Bus. Res., vol. 122, pp. 915–925, Jan. 2021.
[6] A. Narayanan and V. Shmatikov, "Robust de-anonymization of large sparse datasets," in Proc. IEEE Symposium on Security and Privacy (S&P 2008), Oakland, California, USA, 2008, pp. 111–125.
[7] Ministry of Internal Affairs and Communications, "Fiscal year 2017 information and communications white paper," 2017. [Online]. Available: https://www.soumu.go.jp/johotsusintokei/whitepaper/h29.html.
[8] European Union, "General data protection regulation," 2016. [Online]. Available: https://gdpr-info.eu/.
[9] X. Yao, F. Farha, R. Li, I. Psychoula, L. Chen, and H. Ning, "Security and privacy issues of physical objects in IoT: Challenges and opportunities," Digit. Commun. Netw., vol. 7, no. 3, pp. 373–384, Aug. 2021.
[10] R.L. Rivest, L. Adleman, and M. L. Dertouzos, "On data banks and privacy homomorphisms," Found. Secure Comput., vol. 4, no. 11, pp. 169–180, 1978.
[11] M. Ben-Or, S. Goldwasser and W. Avi, "Completeness theorems for non-cryptographic fault-tolerant distributed computation," in Proc. of the 20th Annual ACM Symposium on Theory of Computing (STOC'88), Chicago, Illinois, USA, 1988, pp. 1–10.

[12] O. Goldreich, S. Mical and A. Wigderson, "How to play ANY mental game," in Proc. of the 19th Annual ACM Symposium on Theory of Computing (STOC'87), New York, New York, USA, 1987, pp. 218–229.

[13] L. Sweeney, "K-anonymity: A model for protecting privacy," Int J Uncertain., Fuzz., vol. 10, no. 5, pp. 557–570, Oct. 2002.

[14] K. LeFevre, D. J. DeWitt and R. Ramakrishnan, "Mondrian multidimensional K-anonymity," in Proc. of the 22nd International Conference on Data Engineering (ICDE'06), Atlanta, GA, USA, 2006, pp. 25–25.

[15] J.-W. Byun, A. Kamra, E. Bertino, and N. Li, "Efficient k-anonymization using clustering techniques," in Proc. of the 12th International Conference on Database Systems for Advanced Applications (DASFAA 2007), Bangkok, Thailand, 2007, pp. 188–200.

[16] G. Loukides,. and JH. Shao, "An efficient clustering algorithm for k-anonymisation," J. Comput. Sci. Technol., vol. 23, no. 2, pp. 188–202, Mar. 2008.

[17] G. Aggarwal, R. Panigrahy, T. Feder, D. Thomas, K. Kenthapadi, S. Khuller, *et al*., "Achieving anonymity via clustering," ACM Trans. Algorithms, vol. 6, no. 3, pp. 49:1–49:19, Jun. 2010.

[18] J.H. Cheon, A. Kim, M. Kim, and Y. Song, "Homomorphic encryption for arithmetic of approximate numbers," in Proc. of the 23rd International Conference on the Theory and Applications of Cryptology and Information Security (ASIACRYPT 2017), Hong Kong, China, 2017, pp. 409–437.

[19] W. Jiang, and C. Clifton, "A secure distributed framework for achieving k-anonymity," VLDB J., vol. 15, no. 4, pp. 316–333, Nov 2006. "Synthetic structure of industrial plastics," in *Plastics,* 2nd ed., vol. 3, J . Peters, E d . New Y o r k , NY, USA: McGraw-Hill, 1964, pp. 15–64.